\documentclass[12pt]{article}
\usepackage{latexsym,graphicx,multirow}
\usepackage{amssymb}
\usepackage{amsmath}
\usepackage{amscd}
\usepackage{amsthm}
\usepackage[left=2cm,top=2.5cm,right=2.5cm,bottom=1.5cm]{geometry}
\usepackage{hyperref}
\usepackage{color}

\usepackage{epstopdf}
\usepackage{cite}
\usepackage{float}
\usepackage[utf8]{inputenc}

\begin{document}

    \begin{center}
        \large{\bf{Kaniadakis Agegraphic Dark Energy}} \\
        \vspace{10mm}
   \normalsize{Suresh Kumar P $^{a,1}$, Bramha Dutta Pandey$^{b,2}$,  Pankaj$^{c,3}$, Umesh Kumar Sharma $^{d,4}$ } \\
    \vspace{5mm}
    \normalsize{$^{a,\:b,\:d}$ Department of Mathematics, Institute of Applied Sciences and Humanities, GLA University
        Mathura-281406, Uttar Pradesh, India}\\
    \vspace{2mm}
    \normalsize{$^{a,\:b}$ IT Department (Math Section), University of Technology and Applied Sciences-Salalah, Oman}\\
    \vspace{2mm}
    \normalsize{$^c$ IT Department (Math Section), University of Technology and Applied Sciences-HCT, Muscat, Oman}\\
    \vspace{2mm}

    $^1$E-mail: sureshharinirahav@gmail.com\\
    $^2$E-mail:bdpandey05@gmail.com\\
    $^3$E-mail: pankaj.fellow@yahoo.co.in \\
    $^4$E-mail: sharma.umesh@gla.ac.in \\
    \vspace{10mm}

     %%%%%%%%%%%%%%%%%%%%%%%%%%%%%%%%%%%%%%%%%%%%%%%%%%%%%%%%%
\end{center}

\begin{abstract}
	In this manuscript, we present a novel dark energy  model to study the nature of dark energy. Non-extensive Kaniadakis entropy and, timescale as infrared cutoff are the major tools of the current study. Age of the Universe will serve the purpose of infrared cutoff. The cosmological characteristics of the proposed dark energy model, as well as the evolution of the cosmos filled with pressure-free matter and the ensuing dark energy candidates, are investigated. The interaction as well as non-interaction among the two sectors will also be considered. The differential equation for the dark energy density parameter, including the expression of the equation of state and deceleration parameters, are derived. The analysis of deceleration parameter clearly shows the universe to transit from decelerated to accelerated phase around $z\approx 0.6$. The squared sound speed is also plotted against redshift $z$ to check the stability behavior of the model for both the cases. \\
\smallskip


\end{abstract}

\section{\textbf{Introduction}}
The dark energy(DE) which is responsible for accelerated expansion \cite{Riess98,Perl99} of the Universe and whose origin is still under investigation, fills about $70\%$ energy content of the cosmos \cite{Peebles87, Ratra87, Wetterich87}. The cosmological constant $\Lambda$ is the most basic approach to the DE problem \cite{Padmanabhan02}. Recent observations motivate an interaction between dark matter (DM) and DE sectors \cite{Amendola99,Zimdahl01, Chimento03, delCampo06, Wang16, Pavon05, Sheykhi11}. Such an interaction also shows that their development is intertwined. As a consequence, the overall energy momentum conservation rule is decomposed as
\begin{equation}
\dot{\rho_m}+3H\rho_m=Q, \label{3.1}
\end{equation}
\begin{equation}
\dot{\rho_d}+3H(1+w_d)\rho_d=-Q, \label{3.2}
\end{equation}
where $\rho_m,\; \rho_d$ and $w_d \left(\equiv \dfrac{P_d}{\rho_d}\right)$ are DM energy density, DE density and equation of state (EoS) parameter respectively with DE pressure $P_d$. $Q>0$ indicates transfer of energy from DM to DE whereas $Q<0$ corresponds for DE to DM transfer.\cite{Wang16} provides detailed reference about interacting DE models. The observational data support $\Lambda-$CDM model \cite{Telkamp18}, but fail to answer cosmic coincidence and fine tuning problems. 
The Agegraphic Dark Energy (ADE) model is a different way of answering the problems above. The ADE model is based on quantum mechanics' uncertainty principle \cite{Cai07}. The space time $t$ uncertainty is given by $\delta t=\beta t_p^{\frac{2}{3}}t^{\frac{1}{3}}$ for Minkowskian space-time with reduced Plank time $t_p$ and dimensionless constant $\beta$ of order unity.

Altering the entropy of the system modifies the corresponding ADE model which signifies the importance of the entropy relation \cite{Cai07, Wei08, Wei09}. Various ADE models are studied in \cite{Wei08a,Sheykhi09, Zadeh20, Sharma20}.
The basic criteria to construct holographic dark energy (HDE) model connects entropy of the system with it's own radius. The Bekenstein-Hawking entropy, which is derived from a black hole and a cosmic application of the Boltzmann-Gibbs entropy, is the most popular. Kaniadakis postulated the Kaniadakis entropy \cite{Kaniadakis02} as a one-parameter generalisation of the Boltzmann-Gibbs entropy, that has been studied in the literature  \cite{Sharma21}. The basic idea behind HDE model is the relation $\rho_dL^4\leq S$, where $L$ represents the longest distance and $S$, the entropy relation used in a black hole of radius $L$ \cite{Li04}. For standard Bekenstein-Hawking entropy, the entropy-area relationship is given by $S_{BH}\propto\dfrac{A}{4G}$ with Newton's constant $G$. The Kaniadakis entropy with dimensionless parameter $K$ is defined as $S_K=\dfrac{1}{K}sinh(KS_BH)$ with $\lim_{K\rightarrow0}S_K=S_{BH}$. Which can be rewritten as $ S_K=S_{BH}+\dfrac{K^2}{6}S_{BH}^3+\mathrm{O(K^4)}$.
  
And hence the inequality $\rho_dL^4\leq S$ leads to \cite{Drepanou21, Pandey21}
\begin{equation}
\rho_d=\dfrac{3c^2M_p^2}{L^2}+K^2M_p^6L^2. \label{3.5}
\end{equation}
In this manuscript we will use (\ref{3.5}) to propose Kaniadakis agegraphic dark energy (KADE) model by employing universe age as IR cutoff and investigate the consequences on the Universe's development. 
The manuscript is organized as follows: Next section is dedicated to the formulation of proposed KADE model for which the universe age is used as the IR cutoff to analyze the universe evolutionary behavior. The investigation is carried out by considering non interaction as well as interaction among the DE and DM sectors of the Universe. We will study the non-interacting KADE model in section 3. One full section is dedicated to study the interaction among two sectors. The final section will summarize the results, conclusion and future possibilities of the model.

\section{\textbf{Formulation of KADE Model}}
Age of the Universe is defined as 
\begin{equation}
T=\int_0^{a(t)} \mathrm{d}t=\int_0^{a(t)}\dfrac{\mathrm{d}a(t)}{Ha(t)}, \label{3.6}
\end{equation}
with scale factor $a(t)$ and the Hubble parameter $H$,  related by $H=\frac{\dot{a}(t)}{a(t)}$. By considering (\ref{3.6}) as IR cutoff, the energy density of KADE defined by (\ref{3.5}) can be written as
\begin{equation}
\rho_d=\dfrac{3c^2M_p^2}{T^2}+K^2M_p^6T^2. \label{3.7}
\end{equation}

The standard ADE model \cite{Cai07} can be recovered for $K=0$. The metric for flat FRW universe is considered as $\mathrm{d}s^2 =-\mathrm{d}t^2 +\delta_{ij} \mathrm{d}x^i  \mathrm{d}x^j a^2(t)$. Assuming the FRW universe to be filled by a pressureless fluid with density $\rho_m$ and KADE density $\rho_d$, the first Friedmann equation is expressed as
\begin{equation}
H^2=\dfrac{1}{3M_p^2}\left(\rho_m+\rho_d\right). \label{3.8}
\end{equation}
Equation (\ref{3.8}) can be rewritten as 
\begin{equation}
\Omega_d+\Omega_m=1,  \label{3.9}
\end{equation}
where $\Omega_m=\dfrac{\rho_m}{3M_p^2H^2},\quad \Omega_d=\dfrac{\rho_d}{3M_p^2H^2}$.\\
Because we want to demonstrate the model's potential, we only chose specific system parameters values, such as $K, c$ and $b^2$, that produce diverse behaviors. We choose $\Omega_{d_0}= 0.70$ and $H(a = 1) = 67.9$ as current values of these parameters.

\section{Non-interacting KADE model}
In this part, we will develop the formulation of  KADE model by considering the independence of DE and DM sectors, i.e. $Q=0$. The expression for EoS parameter is obtained by the time derivative of equation (\ref{3.7}) and using (\ref{3.2}), i.e.
\begin{equation}
w_d=-1-\dfrac{2}{3HT}\left(\dfrac{K^2M_p^4T^4-3c^2}{K^2M_p^4T^4+3c^2}\right), \label{3.10}
\end{equation}
where 
\begin{equation}
T=\sqrt{\dfrac{3H^2 \Omega_d-\sqrt{9H^4\Omega_d^2-12c^2K^2M_p^4}}{2K^2M_p^4}}. \label{3.11}
\end{equation}

Differentiating (\ref{3.8}) w.r.t. time, using (\ref{3.1}) and (\ref{3.2}) we get
\begin{equation}
\dfrac{\dot{H}}{H^2}=-\dfrac{3}{2}\left(1-\Omega_d\right)+\dfrac{\Omega_d}{HT}\left(\dfrac{K^2M_p^4T^4-3c^2}{K^2M_p^4T^4+3c^2}\right).  \label{3.12}
\end{equation}
From (\ref{3.12}), the expression for deceleration parameter $q$ is given by
\begin{equation}
q=-1-\dfrac{\dot{H}}{H^2}=\dfrac{1}{2}(1-3\Omega_d)-\dfrac{\Omega_d}{HT}\left(\dfrac{K^2M_p^4T^4-3c^2}{K^2M_p^4T^4+3c^2}\right). \label{3.13}
\end{equation}
The cosmic time derivative of the KADE density parameter is given by 
\begin{equation}
\dot{\Omega_d}=2H\left(1+q \right))\Omega_d+\dfrac{2\Omega_d}{T}\left(\dfrac{K^2M_p^4T^4-3c^2}{K^2M_p^4T^4+3c^2}\right). \label{3.14}
\end{equation}

In order to analyze the stability of the non-interacting KADE model, the squared sound speed $v_s^2$ expressed by 
\begin{equation}v_s^2=\dfrac{\mathrm{d}P_d}{\mathrm{d}\rho_d}=\dfrac{\dot{P_d}}{\dot{\rho_d}}=\dfrac{\rho_d}{\dot{\rho_d}}\dot{\omega_d}+\omega_d, \label{3.15}
\end{equation}
is studied.

\begin{figure}[H]
\centering
\includegraphics[width=16cm,height=8cm, angle=0]{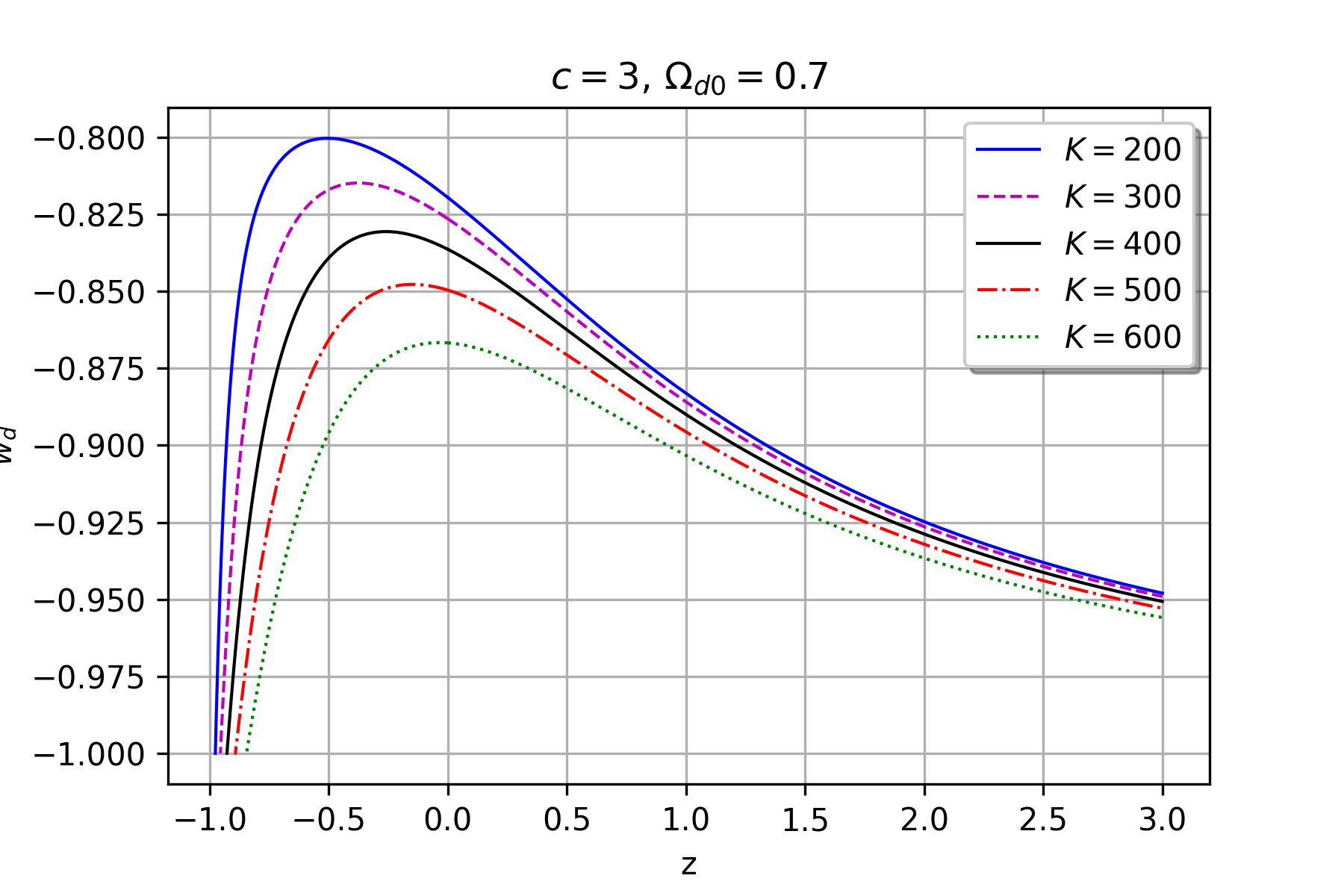}

\caption{\begin{small}
Behavior of $w_d$ vs $z$ for $H_0=67.8, \quad M_p^2=1$ and prescribed values of $K$.
\end{small}}
\label{P3.1} 
\end{figure}

\begin{figure}[H]
\centering
\includegraphics[width=16cm,height=8cm, angle=0]{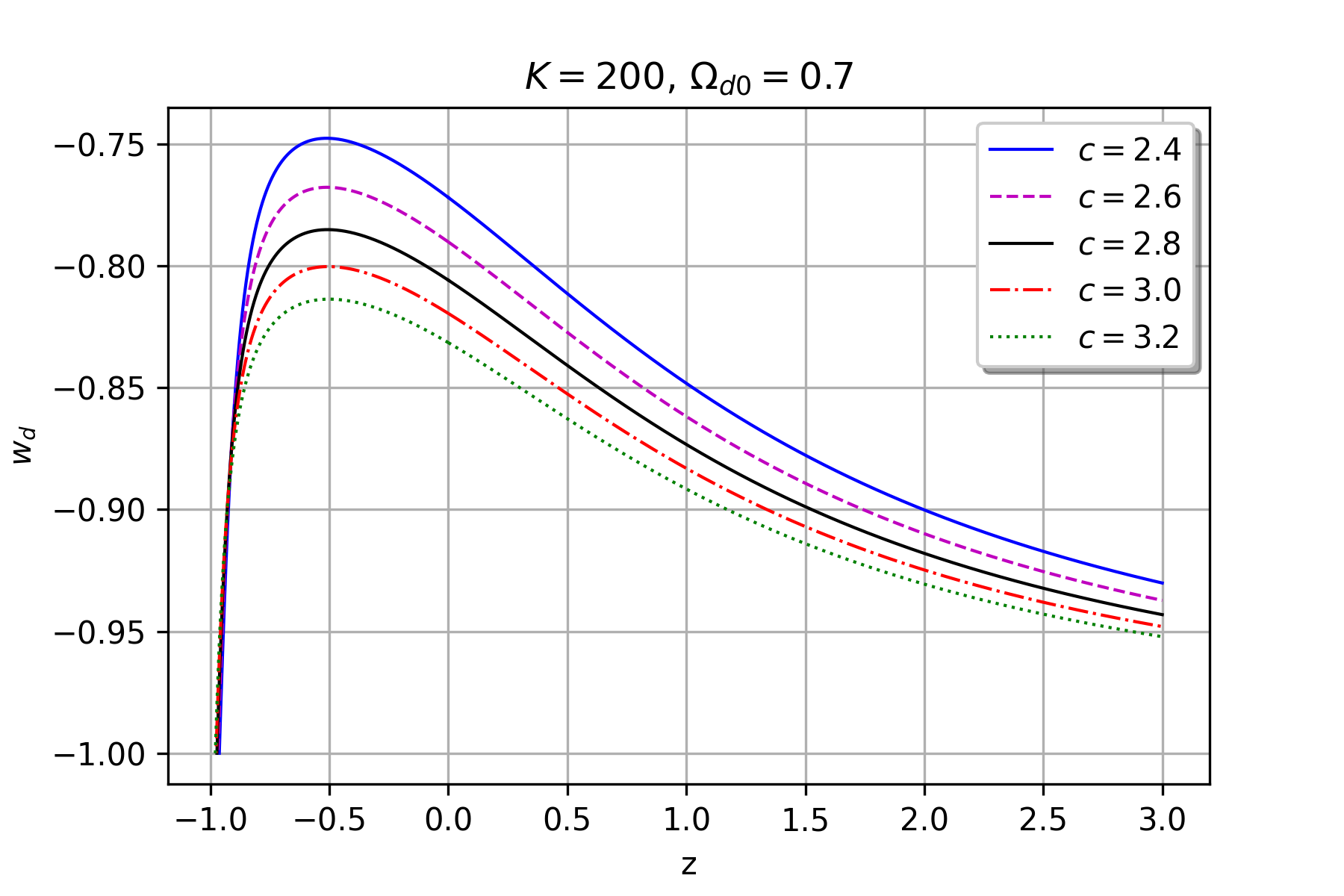}

\caption{\begin{small}
Behavior of $w_d$ vs $z$ for $H_0=67.8, \quad M_p^2=1$ and prescribed values of $c$.
\end{small}}
\label{P3.2} 
\end{figure}

Figures \ref{P3.1} and \ref{P3.2} depict the EoS parameter behavior for redshift $z$. In figure \ref{P3.1}, $c=3$ is fixed and $K$ made to vary from 200 to 600. Whereas, figure \ref{P3.2} is plotted for fixed $K=200$ and varying $c$ from 2.4 to 3.2. In both cases, the model reflects a pure quintessence behavior. All the above plots meet in $w_d=-1$, the $\Lambda-$CDM model in far future. 

\begin{figure}[H]
\centering
\includegraphics[width=16cm,height=8cm, angle=0]{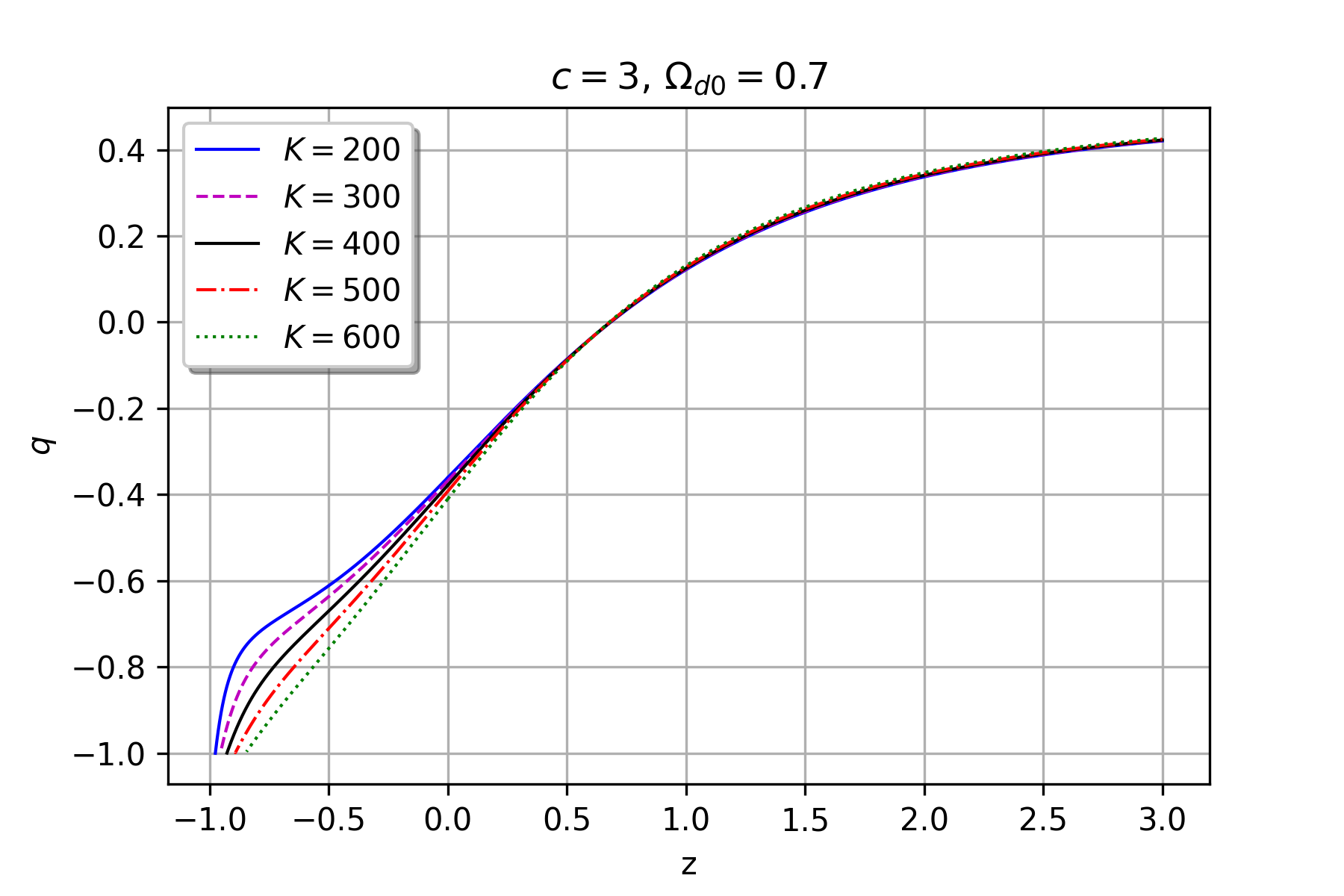}

\caption{\begin{small}
Behavior of $q$ vs $z$ for $H_0=67.8, \quad M_p^2=1$ and prescribed values of $K$.
\end{small}}
\label{P3.3} 
\end{figure}

\begin{figure}[H]
\centering
\includegraphics[width=16cm,height=8cm, angle=0]{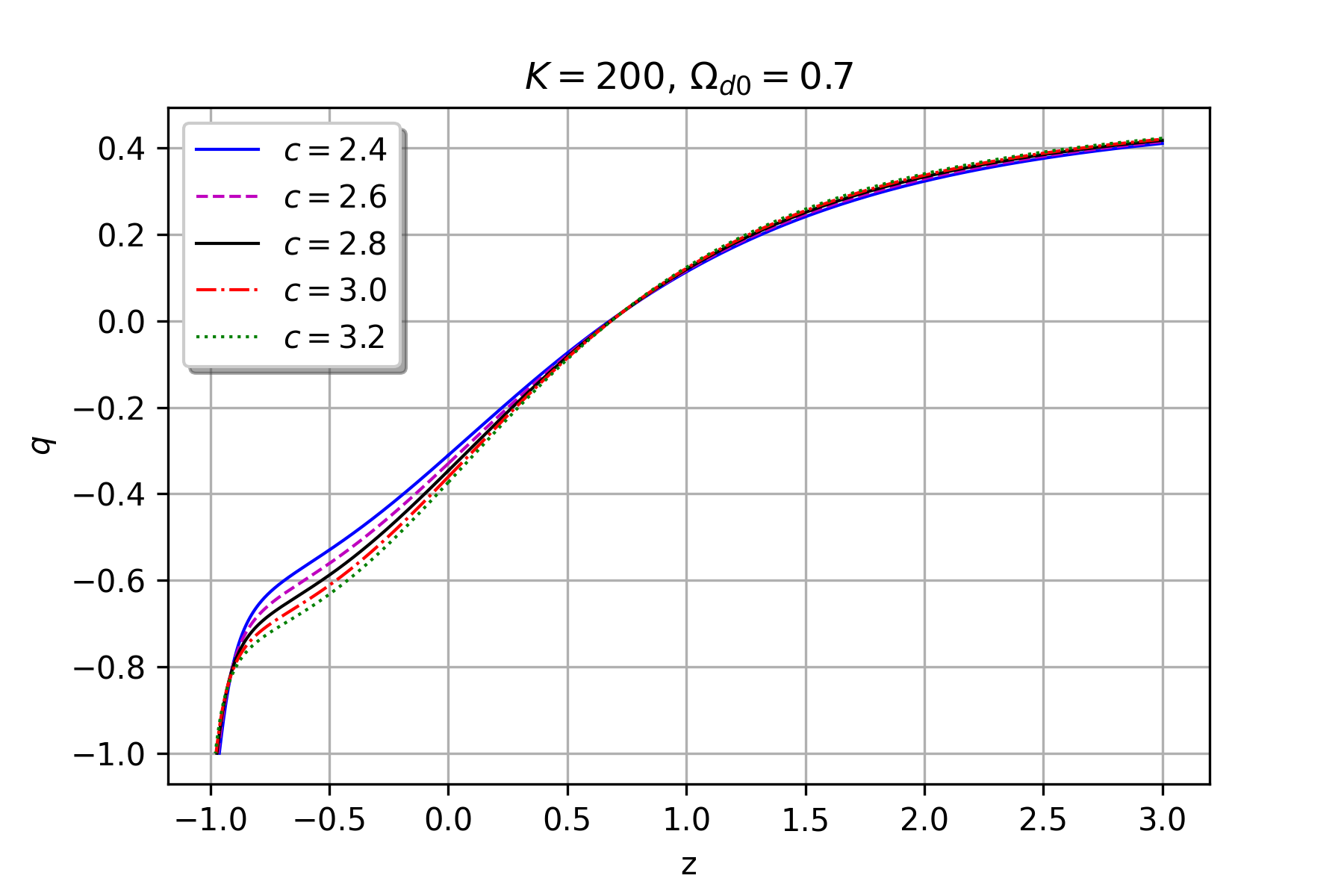}

\caption{\begin{small}
Behavior of $q$ vs $z$ for $H_0=67.8,  \quad M_p^2=1$ and prescribed values of $c$.
\end{small}}
\label{P3.4} 
\end{figure}

The deceleration parameter behavior in figures \ref{P3.3} and \ref{P3.4} clearly says the universe to transit from decelerated to accelerated phase at $z\approx0.6$. The current value of deceleration parameter for considered values of $K$ and $c$ is found to be in the vicinity of $-0.4$.

\begin{figure}[H]
\centering
\includegraphics[width=16cm,height=8cm, angle=0]{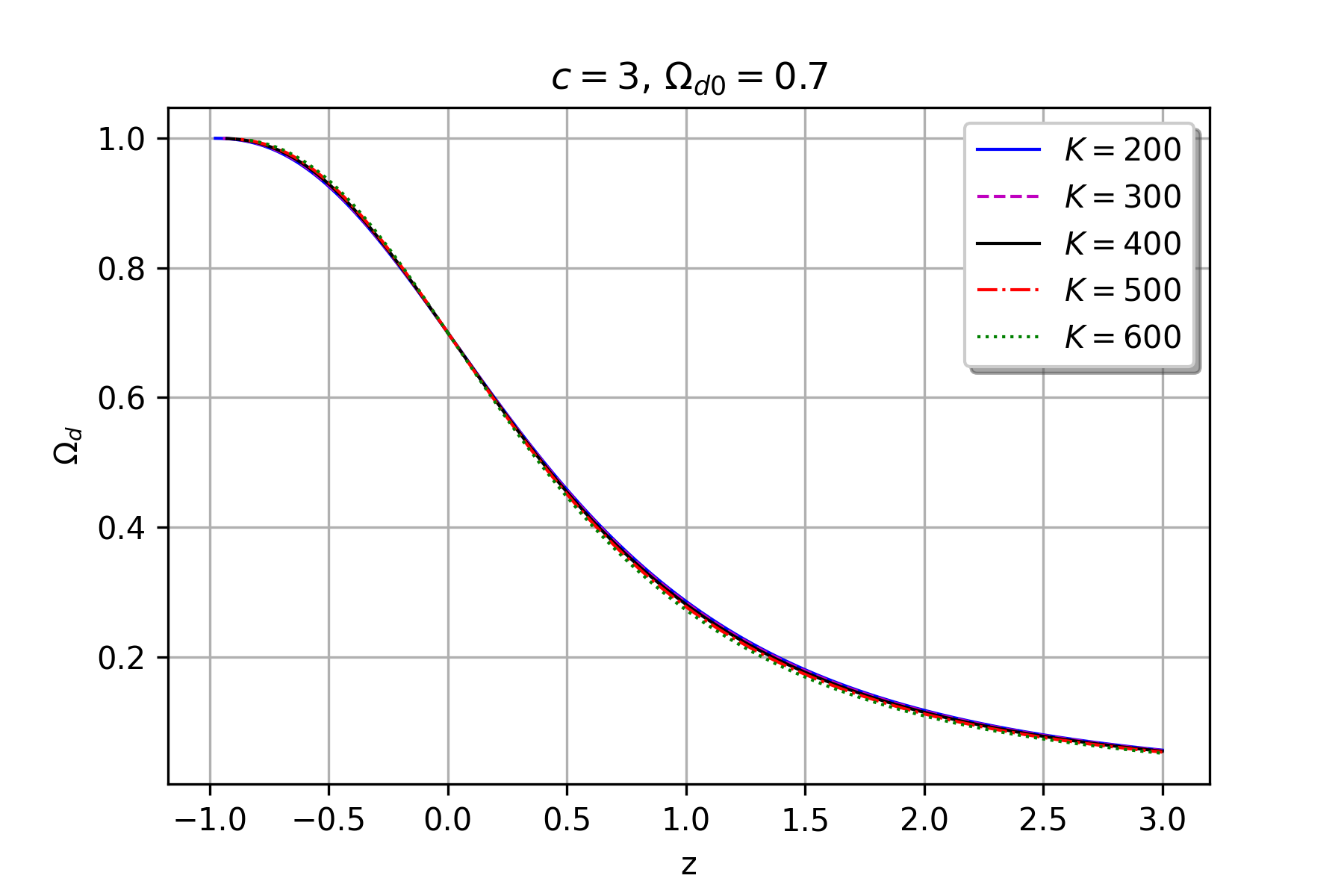}

\caption{\begin{small}
Behavior of $\Omega_d$ vs $z$ for $H_0=67.8,  \quad M_p^2=1$ and prescribed values of $K$.
\end{small}}
\label{P3.5} 
\end{figure}

\begin{figure}[H]
\centering
\includegraphics[width=16cm,height=8cm, angle=0]{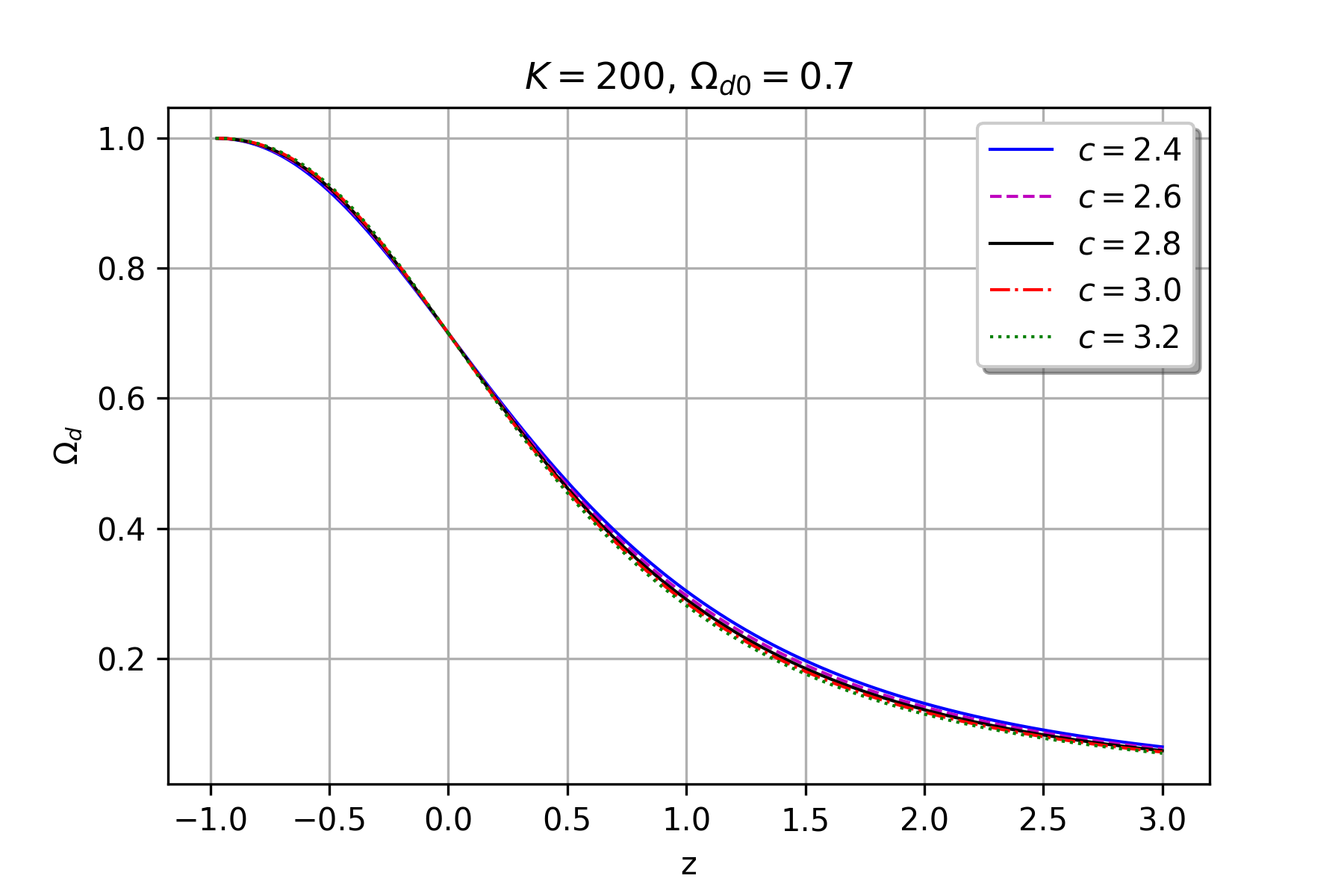}

\caption{\begin{small}
Behavior of $\Omega_d$ vs $z$ for $H_0=67.8, \quad M_p^2=1$ and prescribed values of $c$.
\end{small}}
\label{P3.6} 
\end{figure}

The behavior of KADE density parameter $\Omega_d$ is shown in figures \ref{P3.5} and \ref{P3.6}. For figure \ref{P3.5}, $K$ is allowed to vary from 200 to 600 by fixing $c(=3)$, whereas \ref{P3.6} is plotted based on varying $c$ from 2.4 to 3.2 and fixed $K(=200)$. Both the figures  clearly depict the evolutionary behavior of the universe by telling matter dominated in the past to fully KADE dominated in future. 

\begin{figure}[H]
\centering
\includegraphics[width=16cm,height=8cm, angle=0]{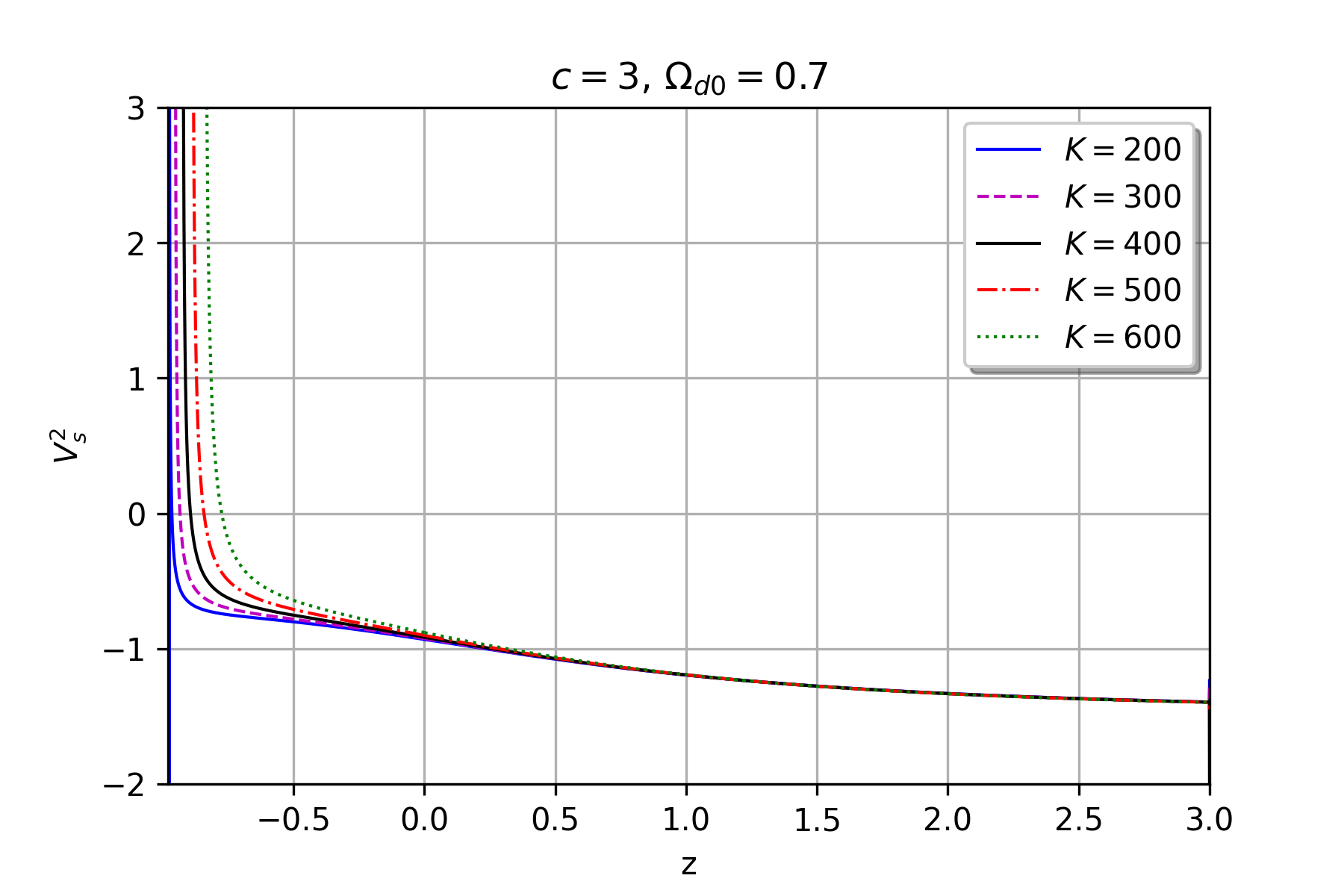}

\caption{\begin{small}
Behavior of $v_s^2$ vs $z$ for $H_0=67.8, \quad M_p^2=1$ and prescribed values of $K$.
\end{small}}
\label{P3.7} 
\end{figure}

\begin{figure}[H]
\centering
\includegraphics[width=16cm,height=8cm, angle=0]{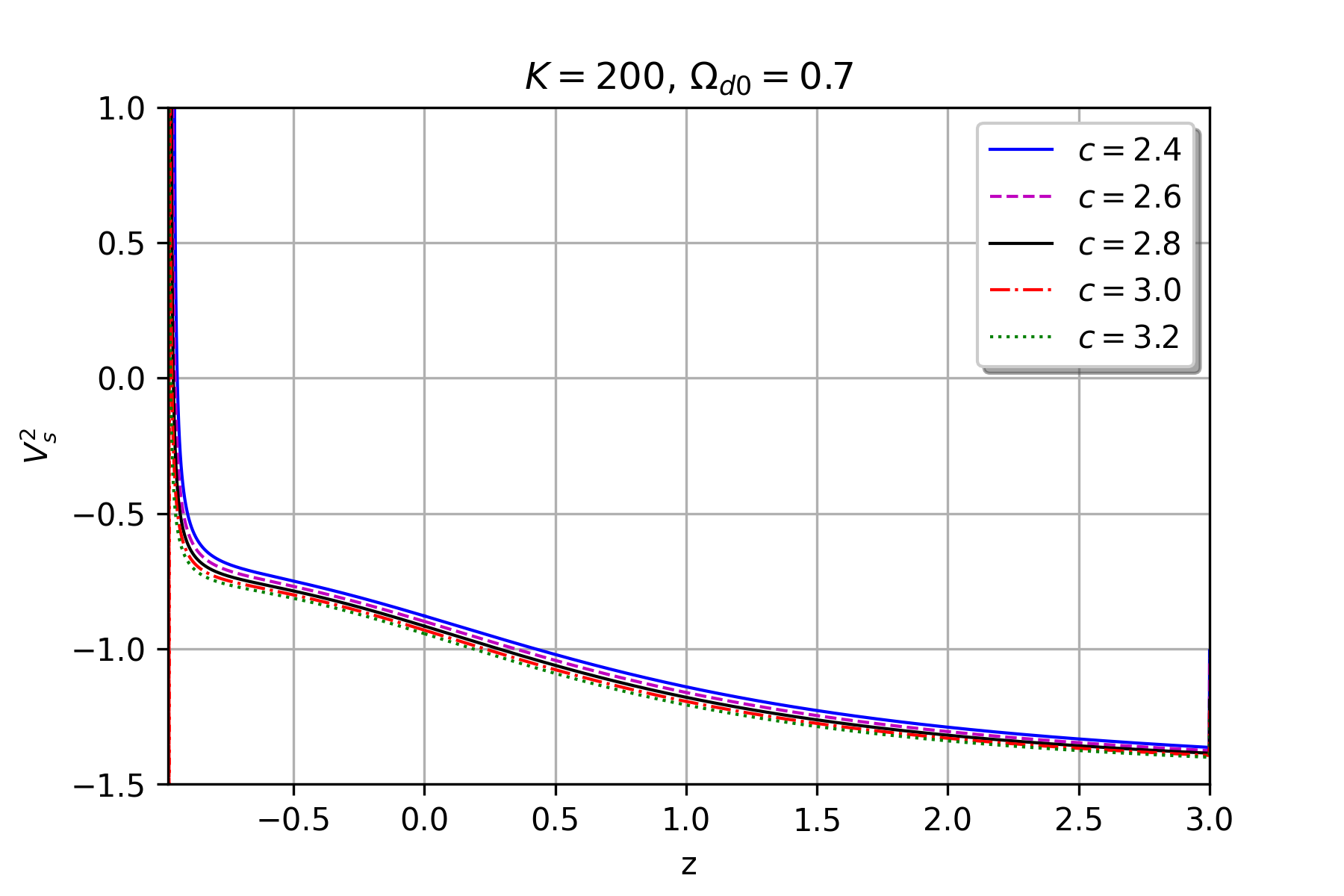}

\caption{\begin{small}
Behavior of $v_s^2$ vs $z$ for $H_0=67.8, \quad M_p^2=1$ and prescribed values of $c$.
\end{small}}
\label{P3.8} 
\end{figure}

The squared sound speed plot in figures \ref{P3.7} and \ref{P3.8} shows the model to be stable for future cosmic behaviors.

\section{Interacting KADE model}
In this section we have considered the interaction between the KADE and the DM sectors coupled by $Q$, defined as
\begin{equation}
Q=3b^2H(\rho_m+\rho_d) \label{3.15a} 
\end{equation}
\cite{Pavon05, Sadeghi13, Honarvaryan15, Zadeh18, Sharma:2019bgp}. The expressions for EoS parameter, deceleration parameter and the KADE density parameter are obtained under such interaction. The stability of the method is analyzed using squared speed of sound. Using time derivative of (\ref{3.7}), making use of (\ref{3.15a}), equation (\ref{3.2}) leads to the expression for EoS parameter
\begin{equation}
w_d=-1-\dfrac{b^2}{\Omega_d}-\dfrac{2}{3HT}\left(\dfrac{K^2M_p^4T^4-3c^2}{K^2M_p^4T^4+3c^2}\right) \label{3.16}
\end{equation}
where $T$ is defined by equation (\ref{3.11}).

From (\ref{3.1}) and (\ref{3.2}) with equation (\ref{3.15a}) and the time derivative of (\ref{3.8}) we get the expressions for deceleration parameter

\begin{equation}
q=\dfrac{1}{2}\left(1-3b^2-3\Omega_d\right)-\dfrac{\Omega_d}{HT}\left(\dfrac{K^2M_p^4T^4-3c^2}{K^2M_p^4T^4+3c^2}\right). \label{3.17}
\end{equation}

On simple adjustment in equation (\ref{3.17}) gives KADE density parameter as
\begin{equation}
\dot{\Omega_d}=2H(1+q)\Omega_d+\dfrac{2\Omega_d}{T}\left(\dfrac{K^2M_p^4T^4-3c^2}{K^2M_p^4T^4+3c^2}\right). \label{3.18}
\end{equation}

\begin{figure}[H]
\centering
\includegraphics[width=16cm,height=8cm, angle=0]{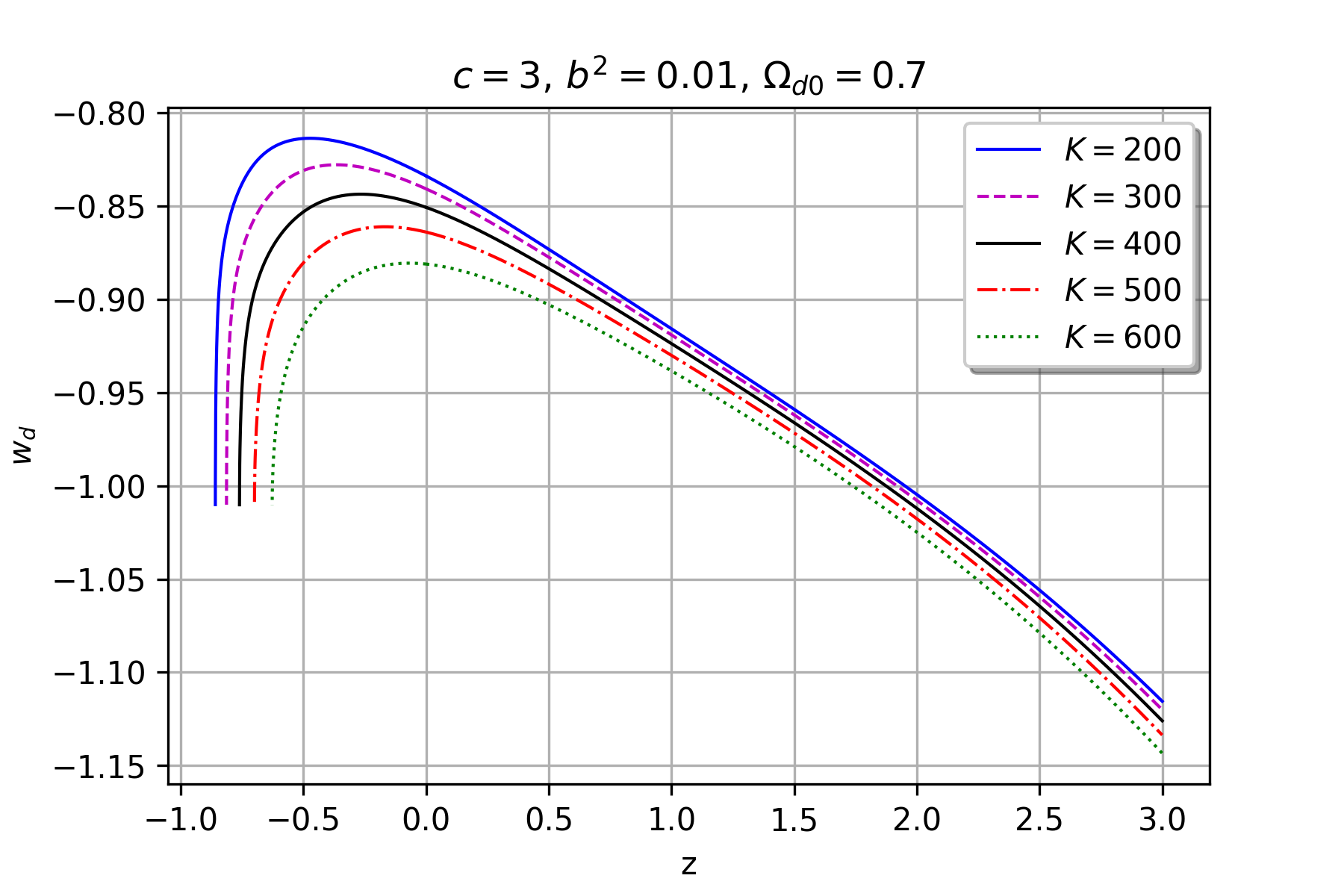}

\caption{\begin{small}
Behavior of $w_d$ vs $z$ for $ H_0=67.8, \quad M_p^2=1$ and prescribed values of $K$.
\end{small}}
\label{P3.9} 
\end{figure}

\begin{figure}[H]
\centering
\includegraphics[width=16cm,height=8cm, angle=0]{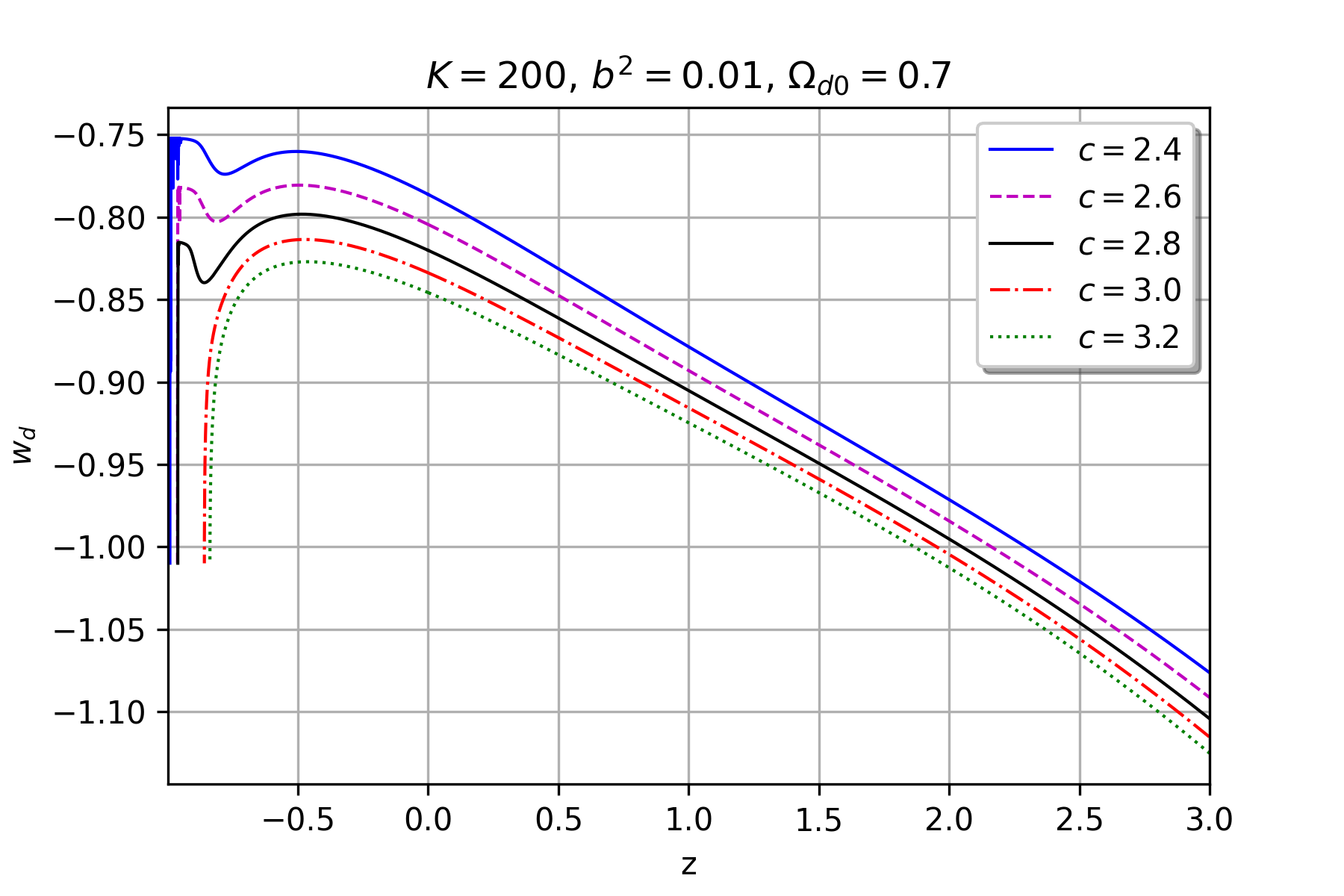}

\caption{\begin{small}
Behavior of $w_d$ vs $z$ for $ H_0=67.8, \quad M_p^2=1$ and prescribed values of $c$.
\end{small}}
\label{P3.10} 
\end{figure}

\begin{figure}[H]
\centering
\includegraphics[width=16cm,height=8cm, angle=0]{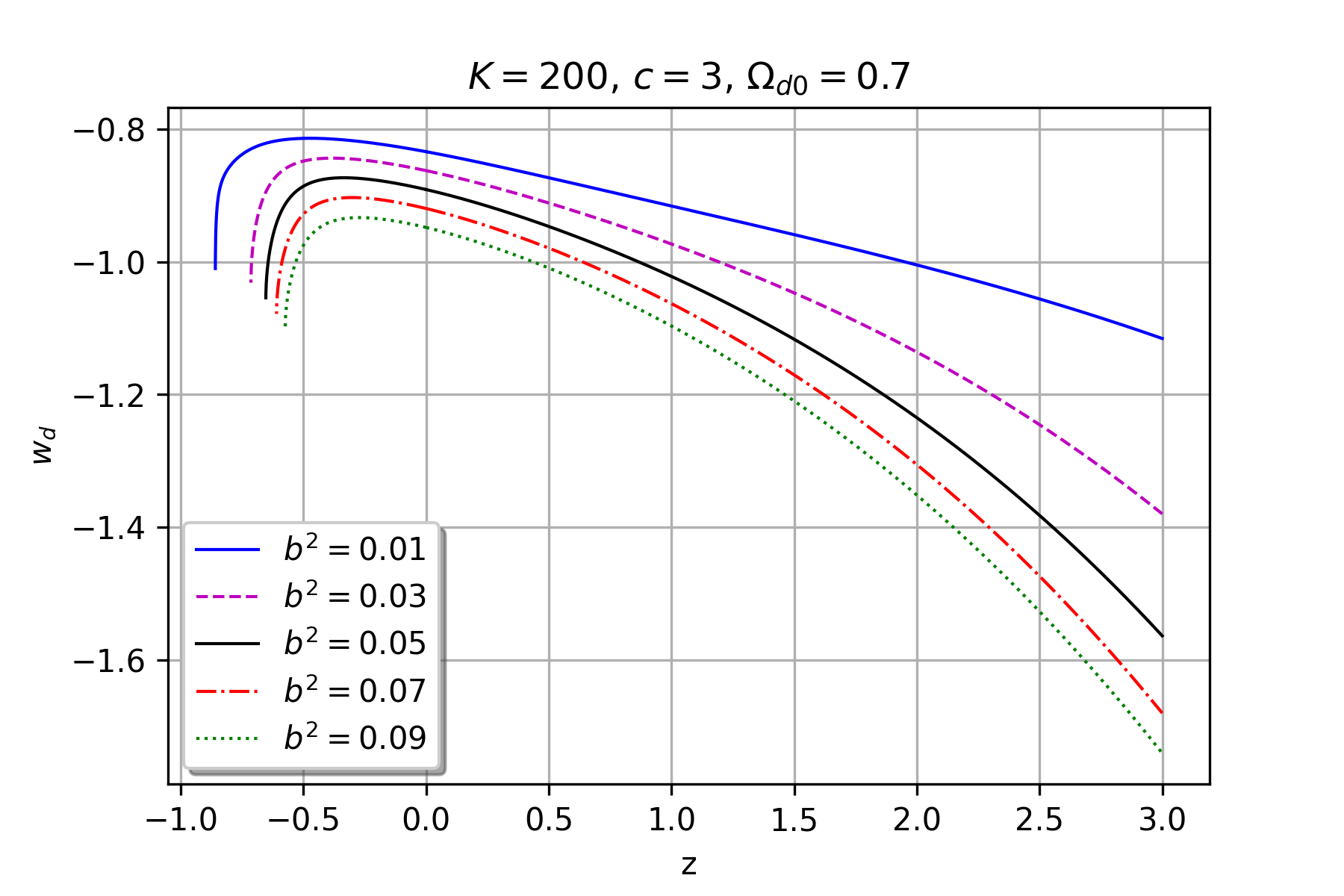}

\caption{\begin{small}
Behavior of $w_d$ vs $z$ for $ H_0=67.8, \quad M_p^2=1$ and prescribed values of $b^2$.
\end{small}}
\label{P3.11} 
\end{figure}

The figures \ref{P3.9} to \ref{P3.11} show the EoS parameter plot against redshift $z$. Clearly, the model describes the universe to evolve from phantom zone. It enters the quintessence zone by crossing the divide line $(w_d=-1)$ and rests the same till present. In far future, again, it enters the phantom zone. 

\begin{figure}[H]
\centering
\includegraphics[width=16cm,height=8cm, angle=0]{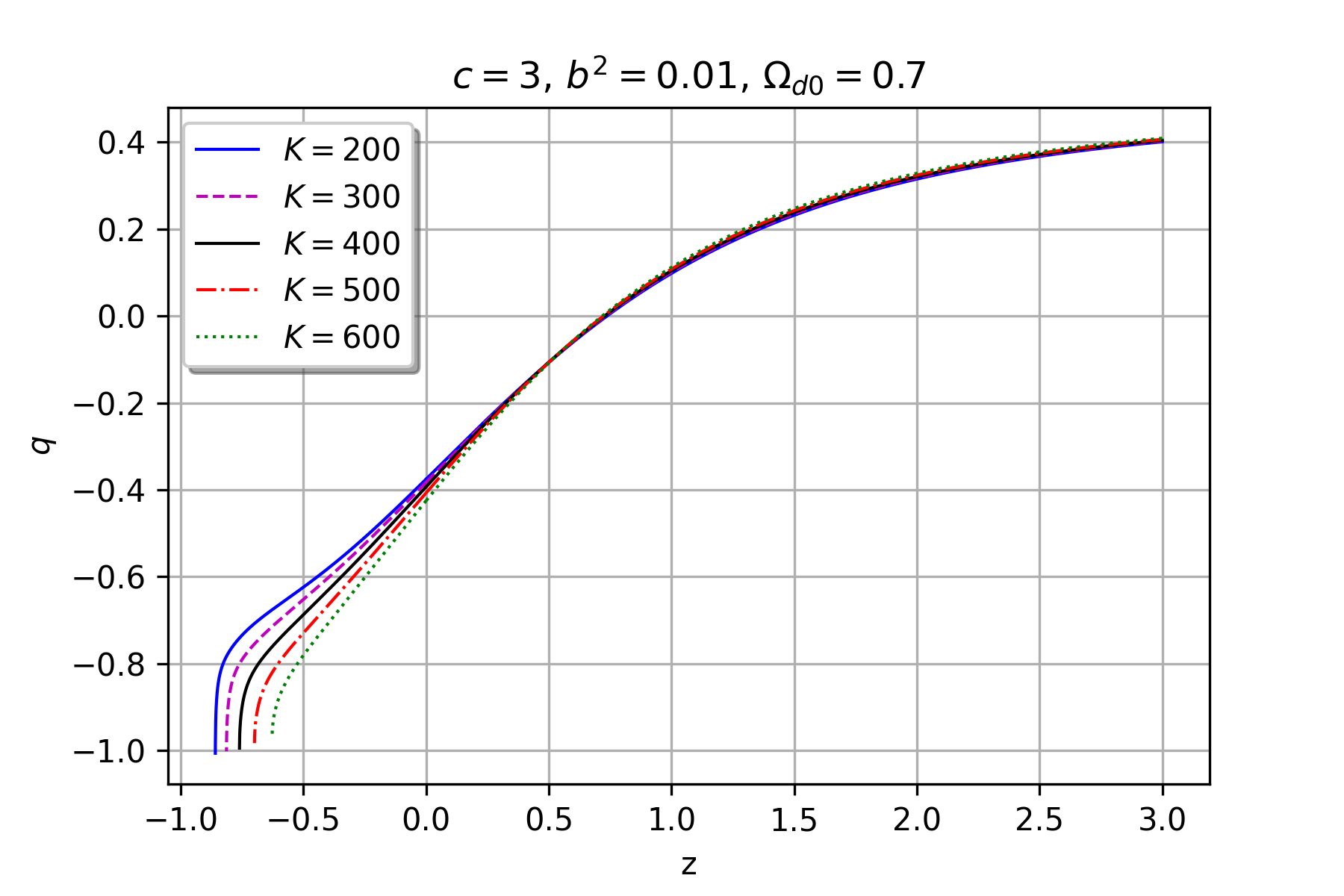}

\caption{\begin{small}
Behavior of $q$ vs $z$ for $ H_0=67.8, \quad M_p^2=1$ and prescribed values of $K$.
\end{small}}
\label{P3.12} 
\end{figure}

\begin{figure}[H]
\centering
\includegraphics[width=16cm,height=8cm, angle=0]{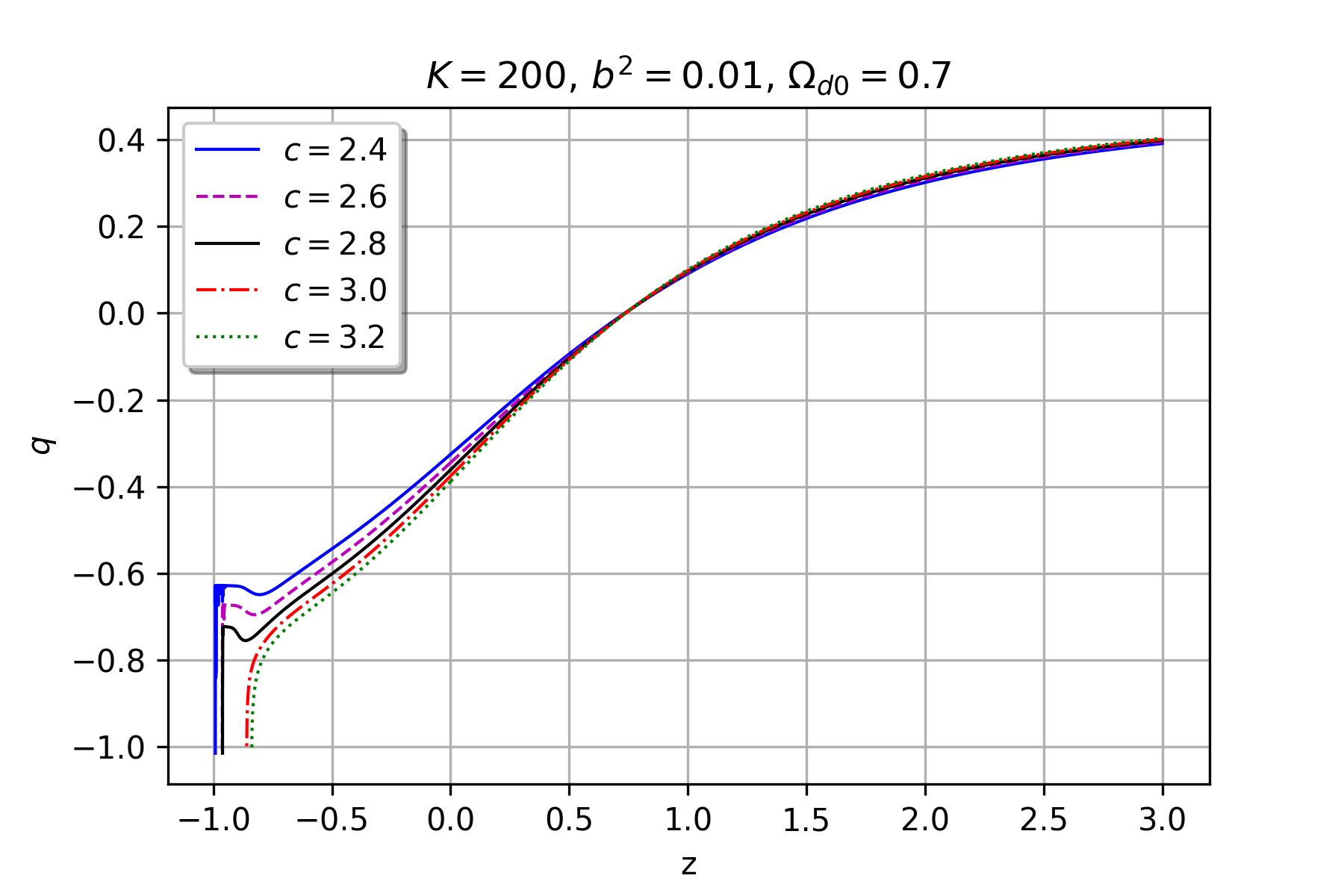}

\caption{\begin{small}
Behavior of $q$ vs $z$ for $ H_0=67.8, \quad M_p^2=1$ and prescribed values of $c$.
\end{small}}
\label{P3.13} 
\end{figure}

\begin{figure}[H]
\centering
\includegraphics[width=16cm,height=8cm, angle=0]{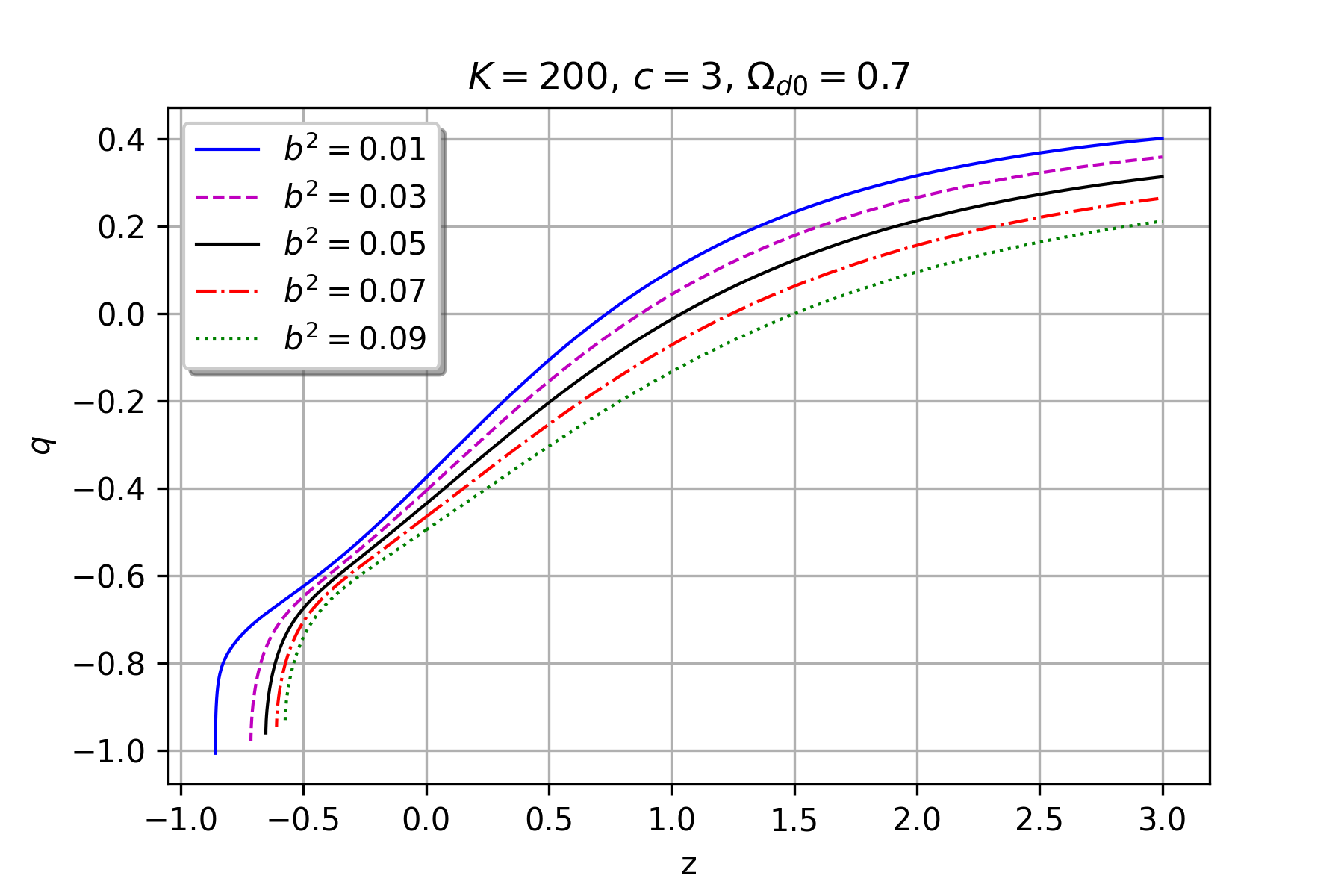}

\caption{\begin{small}
Behavior of $q$ vs $z$ for $ H_0=67.8, \quad M_p^2=1$ and prescribed values of $b^2$.
\end{small}}
\label{P3.14} 
\end{figure}

Figures \ref{P3.12} and \ref{P3.13} are plotted for fixed $b^2$. We observe the universe to enter the accelerated phase some where in $0.6<z<0.8$ and the current value of deceleration parameter $q(z=0)=q_0$ lies in the vicinity of $-0.4$ $(\geq -0.4)$. But, while varying $b^2$ in figure \ref{P3.14} the $z$ interval  become wider. $b^2=0.01$ and 0.03 favors the range $0.6<z<0.8$ but values beyond 0.03 indicate the universe to transit for $z>1$. Such a variation in $b^2$ indicate $q_0$ to be $\leq -0.4$.

\begin{figure}[H]
\centering
\includegraphics[width=16cm,height=8cm, angle=0]{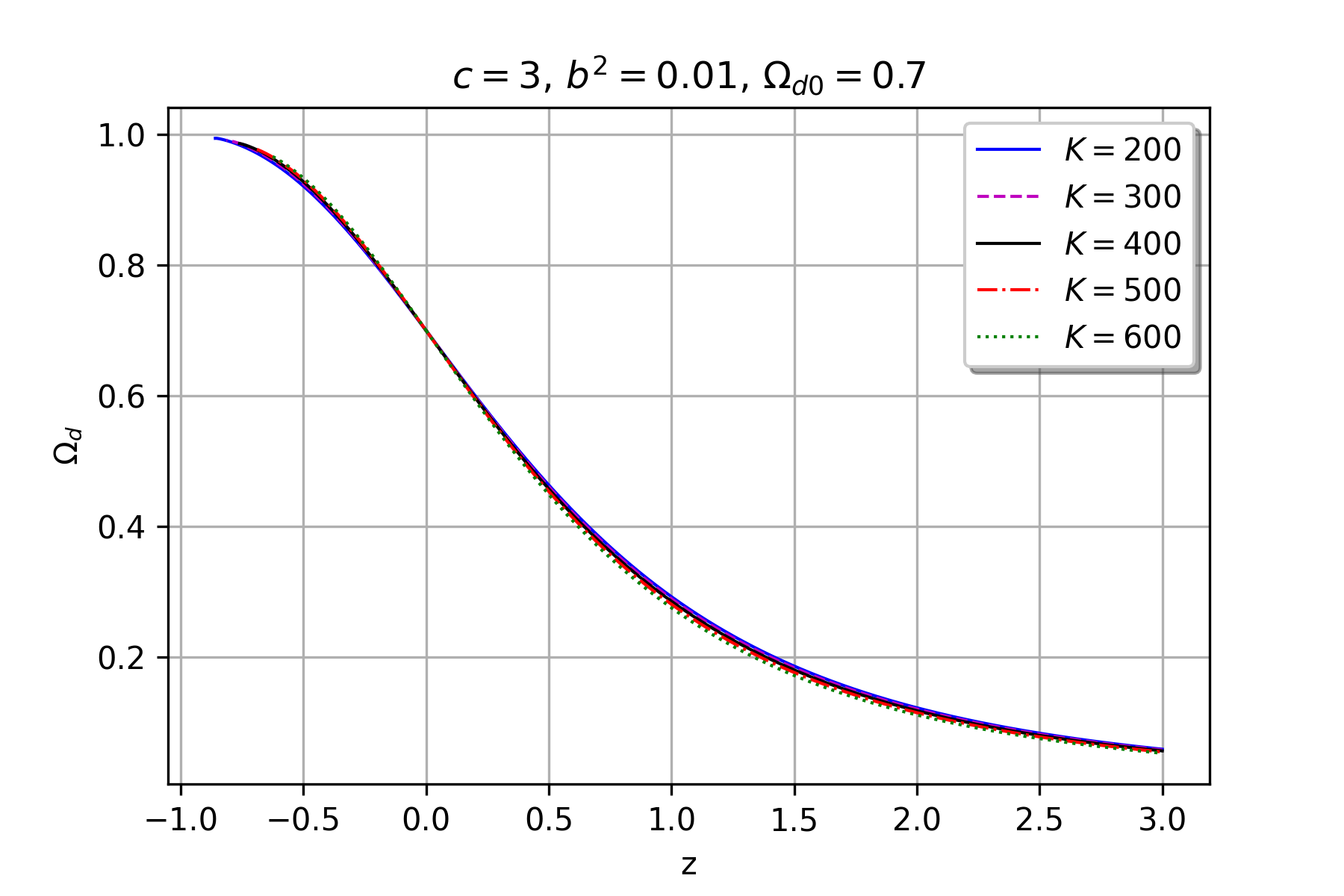}

\caption{\begin{small}
Behavior of $\Omega_d$ vs $z$ for $  H_0=67.8, \quad M_p^2=1$ and prescribed values of $K$.
\end{small}}
\label{P3.15} 
\end{figure}

\begin{figure}[H]
\centering
\includegraphics[width=16cm,height=8cm, angle=0]{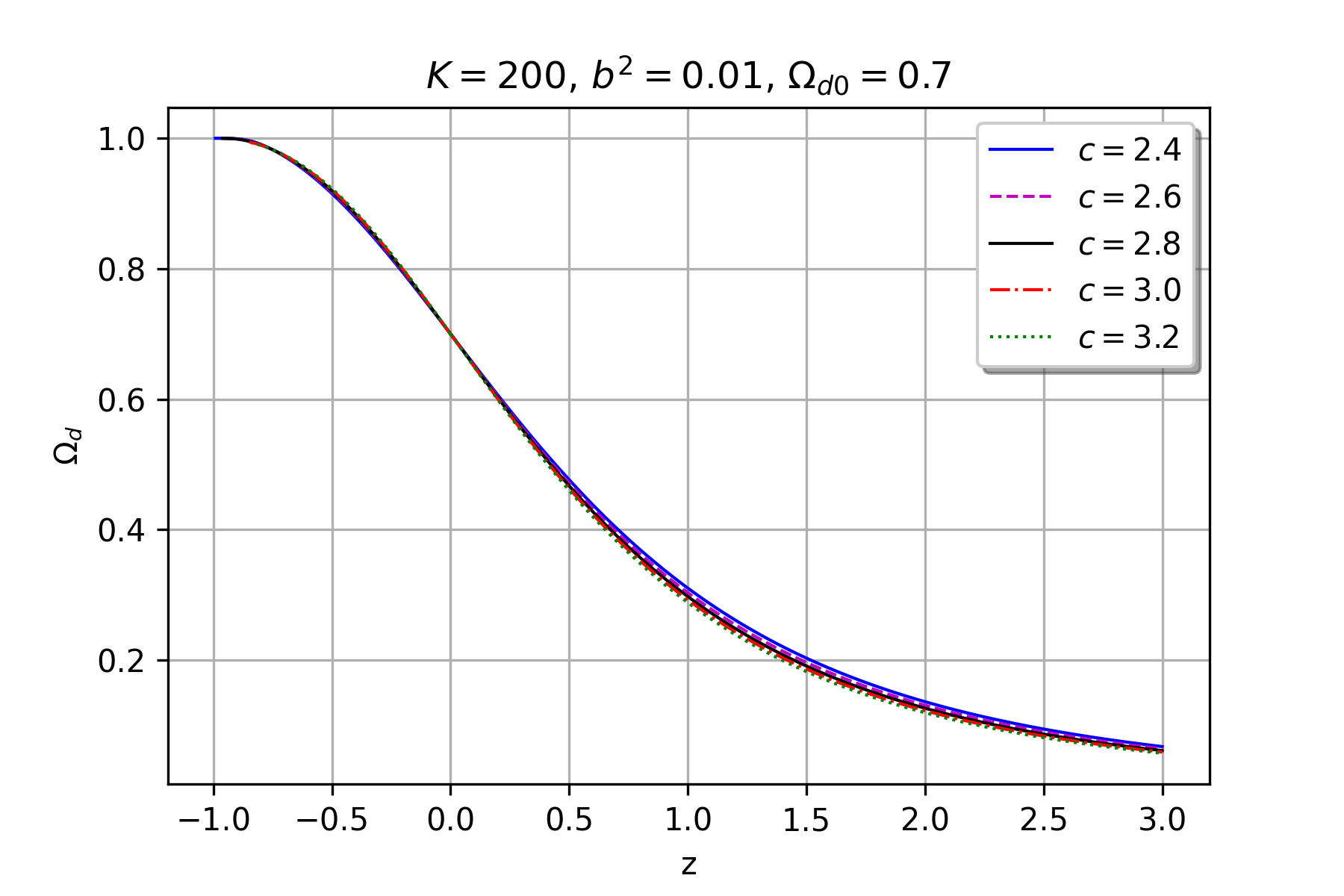}

\caption{\begin{small}
Behavior of $\Omega_d$ vs $z$ for $ H_0=67.8, \quad M_p^2=1$ and prescribed values of $c$.
\end{small}}
\label{P3.16} 
\end{figure}

\begin{figure}[H]
\centering
\includegraphics[width=16cm,height=8cm, angle=0]{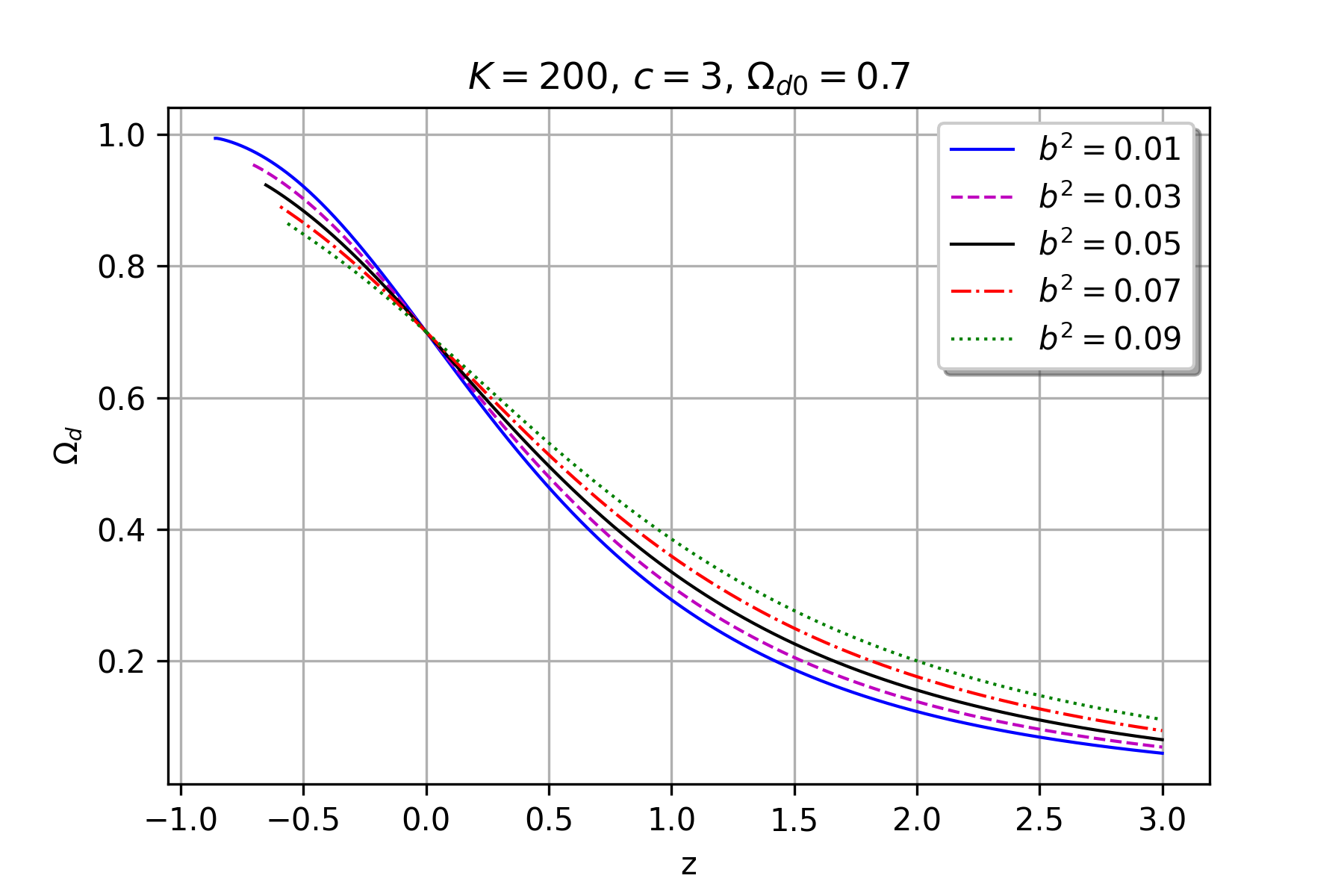}

\caption{\begin{small}
Behavior of $\Omega_d$ vs $z$ for $H_0=67.8, \quad M_p^2=1$ and prescribed values of $b^2$.
\end{small}}
\label{P3.17} 
\end{figure}

Figures \ref{P3.15} and \ref{P3.16} says the universe to be dominated by matter sector in the past i.e. for $z \gtrsim 0.5$ which lies in line with \cite{Frieman08} but increasing values of $b^2$ in figure \ref{P3.17} violate $z \gtrsim 0.5$. All the three figures indicate the universe to be fully dominated by KADE in far future. 

\begin{figure}[H]
\centering
\includegraphics[width=16cm,height=8cm, angle=0]{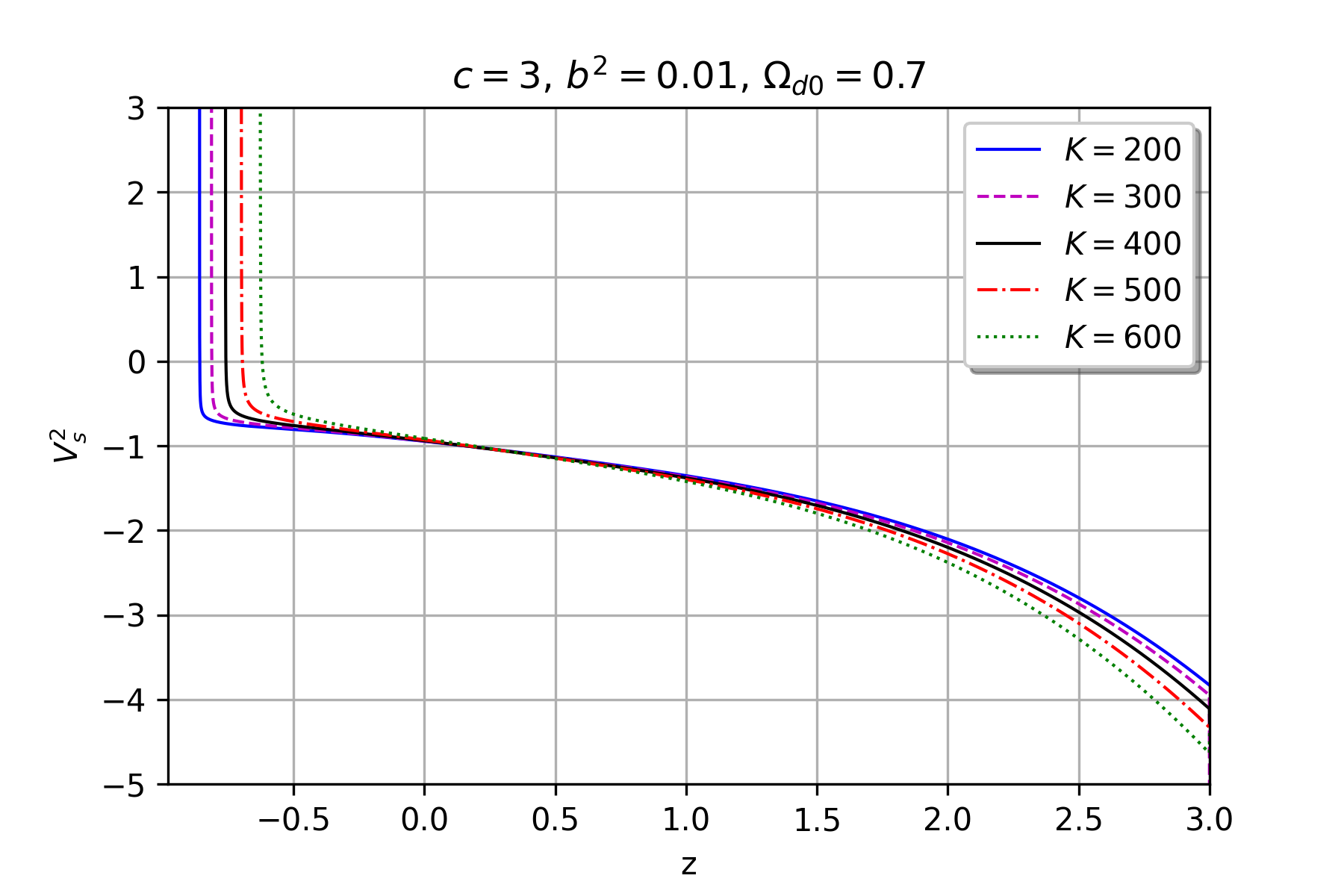}

\caption{\begin{small}
Behavior of $v_s^2$ vs $z$ for $ H_0=67.8, \quad M_p^2=1$ and prescribed values of $K$.
\end{small}}
\label{P3.18} 
\end{figure}

\begin{figure}[H]
\centering
\includegraphics[width=16cm,height=8cm, angle=0]{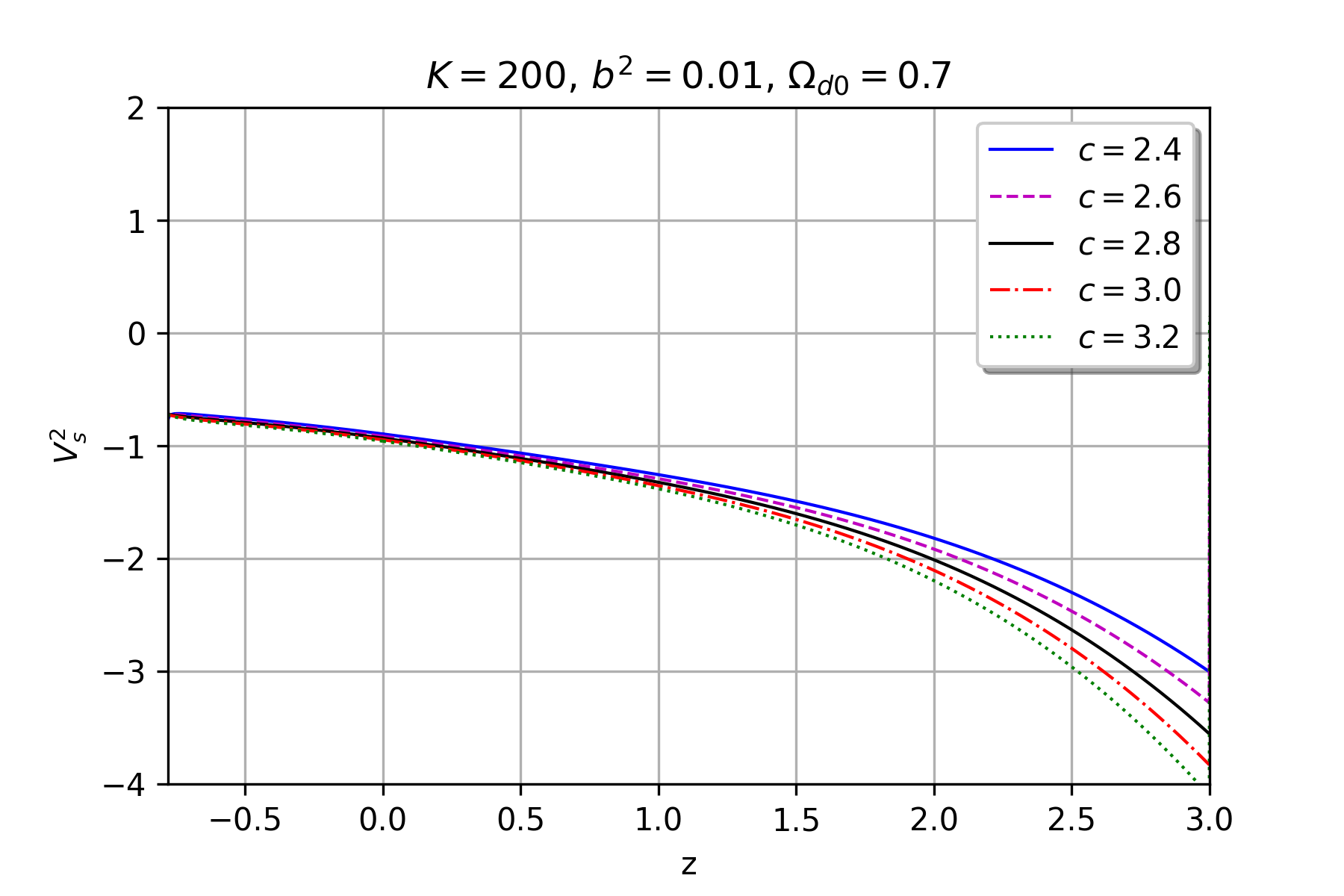}

\caption{\begin{small}
Behavior of $v_s^2$ vs $z$ for $H_0=67.8, \quad M_p^2=1$ and prescribed values of $c$.
\end{small}}
\label{P3.19} 
\end{figure}

\begin{figure}[H]
\centering
\includegraphics[width=16cm,height=8cm, angle=0]{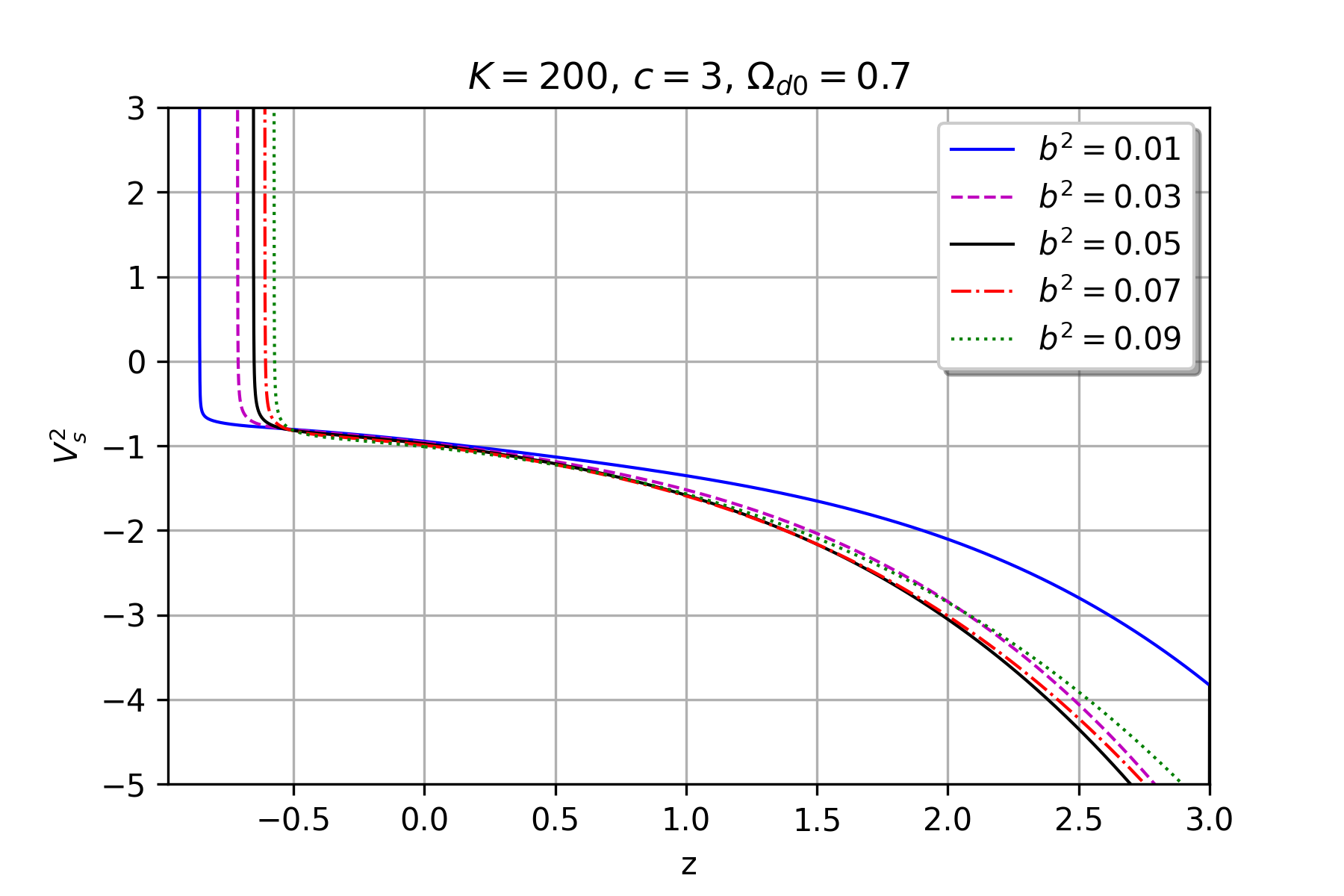}

\caption{\begin{small}
Behavior of $v_s^2$ vs $z$ for $ H_0=67.8, \quad M_p^2=1$ and prescribed values of $b^2$.
\end{small}}
\label{P3.20} 
\end{figure}

The $v_s^2$ behavior in figures \ref{P3.18} and \ref{P3.20}, which is based on variation either in $K$ or $b^2$, reflects the model to be stable in future predictions. While varying $c$ indicates the model to be unstable as evidenced in figure \ref{P3.19}.

\section{Conclusion}

In this study, age of the universe is considered as the IR cut-off and the entropy role is played by Kaniadakis entropy. The evolutionary behavior of the universe is studied by considering the matter and energy sectors to independent as well as interacting. If we consider the two sectors to be fully independent, the KADE model becomes a pure quintessence model as evidenced from figure \ref{P3.1} and \ref{P3.2}. On considering the interaction among the two sectors, figures \ref{P3.9} to \ref{P3.11} shows that the universe evolved from phantom zone and enters to quintessence region by crossing the divide line $(w_d=-1)$ and rests there at present but again enters to phantom zone in far future. Hence the KADE model behaves phantom like in past and far future, whereas quintessence like at present. This reflects the richness of the KADE model. A close look at figures \ref{P3.3}, \ref{P3.4}, \ref{P3.12} and \ref{P3.13} indicates the transition of the universe to accelerated phase happen in $0.6 \leq z \leq 0.8$ irrespective of the interaction between the two sectors and the current $q$ value is found to be greater than $-0.4$. But, while varying $b^2$ in figure \ref{P3.14} the value of $q_0$ lies $\lesssim -0.4$ and the interval of transition also become wider. The universe, ruled by matter sector in the past $(z \gtrsim 0.5)$ and completely dominated by KADE in future, is evidenced from figures \ref{P3.5}, \ref{P3.6} and \ref{P3.15}, \ref{P3.16} and hence the domination by KADE happens in the neighborhood of $z=0.5$. The variation in $b^2$ in figure \ref{P3.17} favors matter domination in the past and KADE domination in the future but creates a distinction for the matter to KADE domination. Hence the condition of $z \approx 0.5$ is no more valid for the domination. The squared speed of sound is plotted in figure 7, 8, 18 and 20 shows the stability of the KADE model in far future for both interacting and non-interacting. Figure 19 shows the instability of the interacting case while varying the value of $c$. Hence we conclude the KADE model will be suitable for the study of the future universe. The parameter estimation with observational data is the future study of this work.

\end{document}